# Flexible Graphene/Carbon Nanotube Electrochemical Double-Layer Capacitors with Ultrahigh Areal Performance

*Valentino Romano,[a],[b] Beatriz Martín-García,[a],† Sebastiano Bellani,[a] Luigi Marasco,[a] Jaya Kumar Panda,[a] Reinier Oropesa-Nuñez,[a],[c] Leyla Najafi,[a] Antonio Esau Del Rio Castillo,[a] Mirko Prato,[d] Elisa Mantero,[a] Vittorio Pellegrini,[a],[c] Giovanna D'Angelo,[b] and Francesco Bonaccorso\*[a],[c]*

[a] V. Romano, Dr. B. Martín-García, Dr. S. Bellani, L. Marasco, Dr. J. Kumar Panda, Dr. R. Oropesa-Nuñez, Dr. L. Najafi, Dr. A. E. Del Rio Castillo, E. Mantero, Dr. Vittorio Pellegrini, Dr. F. Bonaccorso
Graphene Labs, Istituto Italiano di Tecnologia, via Morego 30, 16163 Genova, Italy
E-mail: francesco.bonaccorso@iit.it

[b] V. Romano, Prof. G. D'Angelo
Dipartimento di Scienze Matematiche ed Informatiche, Scienze Fisiche e Scienze della Terra, Università di Messina, Viale F. Stagno d'Alcontres 31, S. Agata, 98166 Messina, Italy

[c] Dr. R. Oropesa-Nuñez, Dr. F. Bonaccorso
BeDimensional Spa., Via Albisola 121, 16163 Genova, Italy

[d] Dr. M. Prato
Materials Characterisation Facility, Istituto Italiano di Tecnologia, via morego 30, 16163 Genova, Italy
† V. Romano and B. Martín-García contributed equally to this work

**Abstract:** The fabrication of electrochemical double-layer capacitors (EDLCs) with high areal capacitance relies on the use of elevated mass loadings of highly porous active materials. Herein, we demonstrate a high-throughput manufacturing of graphene/nanotubes hybrid EDLCs. Wet-jet milling (WJM) method is exploited to exfoliate the graphite into single/few-layer graphene flakes (WJM–G) in industrial volume (production rate ~0.5 kg/day). Commercial single/double walled carbon nanotubes (SDWCNTs) are mixed with graphene flakes in order to act as spacers between the graphene flakes during their film formation. The latter is obtained by one-step vacuum filtration, resulting in self-standing, metal- and binder-free flexible EDLC electrodes with high active material mass loadings up to ~30 mg cm$^{-2}$. The corresponding symmetric WJM–G/SDWCNTs EDLCs exhibit electrode energy densities of 539 µWh cm$^{-2}$ at 1.3 mW cm$^{-2}$ and operating power densities up to 532 mW cm$^{-2}$ (outperforming most of the EDLC technologies). The EDCLs show excellent cycling stability and outstanding flexibility even under highly folded states (up to 180°). The combination of industrial-like production of active materials, simplified manufacturing of EDLC electrodes, and ultrahigh areal performance of the as-produced EDLCs are promising for novel advanced EDLCs designs.

**Introduction**

Electrochemical double-layer capacitors (EDLCs) store energy electrostatically at the interface between a high-surface area electrode and an electrolyte.[1,2] The intrinsic reversibility of ion adsorption processes onto highly porous electrodes and the absence of electrochemical reactions converting electrical energy in chemical form, otherwise present in lithium-ion batteries,[1–3] determines the main properties of EDLCs: fast charge rates (from milliseconds to seconds)[4] or high specific power (>10$^3$ W kg$^{-1}$),[4] specific energy in the order of 10 Wh kg$^{-1}$,[4] and long life stability (millions of charge/discharge cycles)[5]. Such features allow EDLCs to bridge the gap between the performance of batteries, typically delivering low maximum specific power (< 10$^3$ W kg$^{-1}$), and the one of conventional capacitor, storing limited specific energy density (< 0.1 Wh kg$^{-1}$).[2,4,5] Consequently, EDLCs suit for several applications, including electric hybrid vehicles, pulse laser technologies, memory back-up in electronic devices[6] and communication systems,[7] and burst-mode power delivery.[5,8–13] Furthermore, the fabrication of flexible electrodes with high areal capacitance (C$_{areal}$), *i.e.*, high energy density, is of paramount importance for application with limited usable area, including portable wearable electronics (available space in the human body is ~1.5 m$^2$)[14–20] as well as micro-electronics.[21–23]

In this context, nanostructured allotropes of carbon, including activated carbons (ACs)[24,25] carbide-derived carbon[26,27], graphene[28,29], carbon nanotubes (CNTs)[30–32], are intensively investigated as EDLC active materials.[1,2,4] In particular, commercial EDLCs make use of ACs[33] due to their large surface area (500 – 3500 m$^2$ g$^{-1}$)[2] and low cost (< 10 USD kg$^{-1}$).[34] However, the presence of micropores (pores whose diameter is < 2 nm)[2,35] in the AC structure limits the accessibility of electrolyte ions, limiting their maximum power density.[36,37] Moreover, the AC-based electrodes are typically produced by depositing an AC-based paste (prepared by mixing ACs with conductivity enhancers and binders in the form of slurry) on a metallic substrate, acting as current collector.[38] By adopting this method, the deposition of active material can reach mass loadings up to ~10 mg cm$^{-2}$. However, higher active material mass loading values result in brittle films, which undermine the mechanical integrity of the final EDLCs as well as light device packaging with flexible polymeric components.[39,40] Therefore, high-areal performance AC-based EDLCs often result in stiff and cumbersome device which are not intended as flexible energy supply.[41]

With the aim to balance the electrochemical/mechanical properties, and the processing technologies during EDLC design, and the processing technologies during EDLC design, a strong research effort has been focused on the search for carbon-based active materials alternative to ACs.[41–46] Among them, graphene is one of the most promising[43,46–49], due to its physical and chemical properties *i.e.*, charge carriers mobility (> 2000 cm$^2$ V$^{-1}$ s$^{-1}$ on SiO$_2$ substrates)[50,51], theoretical specific surface area (~2600 m$^2$ g$^{-1}$)[52] and excellent mechanical properties[53] (Young module of ~1 TPa)[54] and elastic limit (~20%).[55] However, on the one hand, graphene flakes tend to re-stack during their deposition due to van der Waals forces,[56–58] forming piles oriented horizontally with respect to the substrates (*i.e.*, the current collectors).[59,60] This reduces the surface area accessible to the electrolyte, especially in EDLCs with sufficient active material mass loading (*i.e.*, 1–10 mg cm$^{-2}$), limiting their areal performance and light packaging in practical devices.[61–63] This drawback has been mitigated by engineering graphene-based structures with different topological complexities, including zero-dimensional (0D) graphene quantum dots,[64,65] one dimensional (1D) graphite nanofibers,[66] three-dimensional (3D) graphene foams[67,68] and graphene-based hydrogels[69,70]. However, the production of such structures typically involves chemical functionalization[71,72] or template-assisted growth,[73,74] which poses severe limitations for the industrial manufacturing of EDLCs.[75]

Carbon nanotubes have been also hybridized with graphene flakes to act as spacers between the flakes,[58,76–78] as well as to anisotropically orient the latter ones, increasing their electrochemically accessible surface area in vertical (*i.e.*, sandwich-like) EDLC configurations.[58,76–78] On the other hand, the scalable production of graphene with high quality is challenging,[79–81] slowing down the manufacturing of graphene-based EDLCs on an industrial-scale.[82,83] One of the most promising scalable methods to produce graphene is the liquid phase exfoliation (LPE) method,[84] *i.e.*, the exfoliation of graphite into single/few-layer graphene (SLG/FLG) flakes in liquid solvents by means of cavitation[85,86] or shear forces.[79,86] The main advantage of LPE, compared to bottom-up approaches,[84] is the possibility to produce and subsequently process SLG/FLG flakes in a liquid phase to obtain functional inks,[87] compatible with solution-processed deposition, *e.g.*, spray-coating[88], vacuum filtration[89], ink-jet printing[90,91] and drop-casting[92]. However, for prototypical LPE methods, *i.e.*, ultrasonication,[85,86,93] and shear mixing,[79,94] the volume of solvent required to produce 1 g of exfoliated material, (V$_{1gram}$), the time required to obtain 1 g of exfoliated material (t$_{1gram}$), as well as the ratio between the weight of the final graphitic material and the weight of the starting graphite flakes (*i.e.*, exfoliation yield –Y$_W$–), are still insufficient for industrial-scale production rate (in the order of kg/day).[93,95,96] Recently, our group presented an innovative method for exfoliating the graphite (as well as other layered crystals) based on high-pressure wet-jet-milling (WJM),[81,97,98] which produces 2 L h$^{-1}$ of highly concentrated (10 g L$^{-1}$) defect-free and high quality SLG/FLG flakes dispersions in N-methyl-2-pyrrolidone (NMP),[81,97] making the scaling-up of two dimensional material more affordable (production rate ~0.5 kg/day).

In this work, we demonstrate that WJM-produced SLG/FLG flakes (WJM–G) are ideal for an industrial-like one-step fabrication of flexible EDLC electrodes through vacuum filtration of WJM–G dispersion. To avoid WJM–G restacking during film deposition, commercial CNTs have been used as spacers between the flakes[58,76], as well as active material.[30,32]

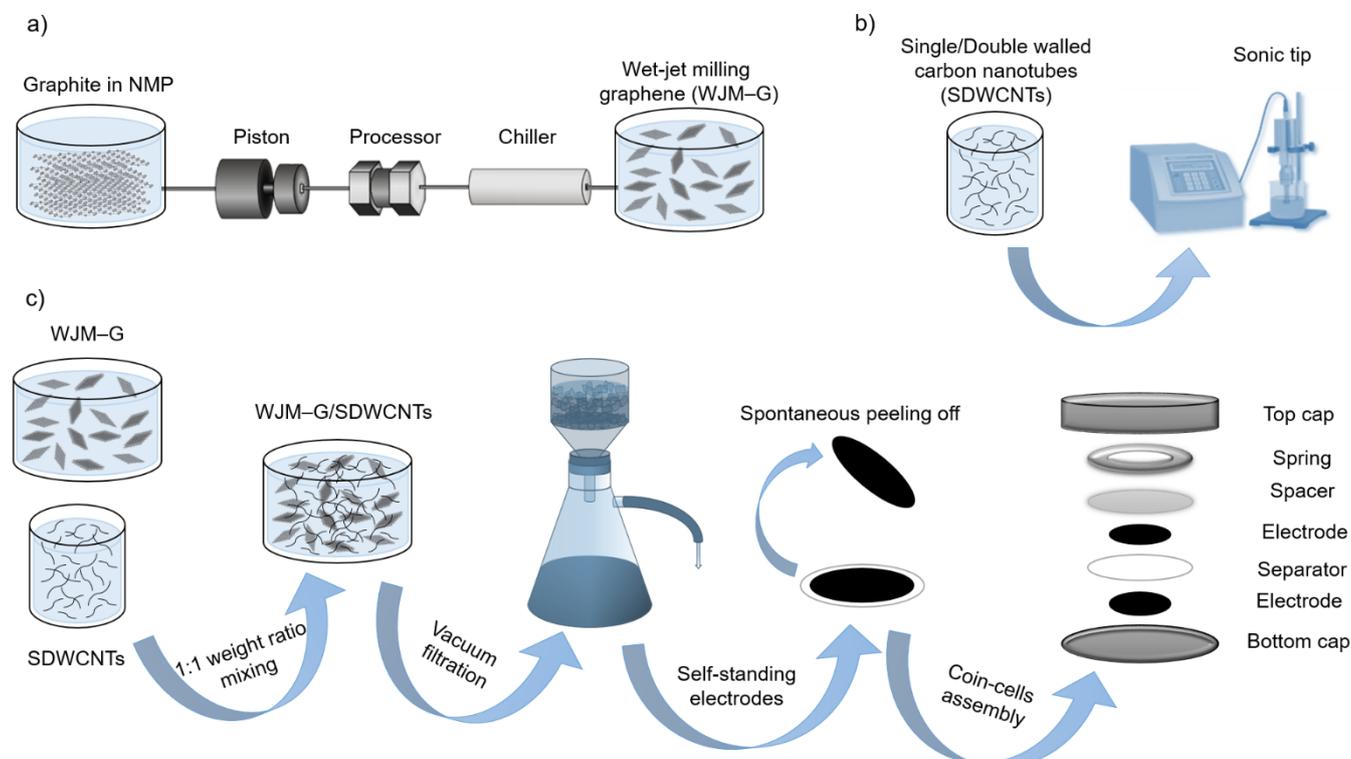

**Fig. 1.** Production of graphene-based materials and manufacturing of EDLCs. (a) Sketch of the WJM exfoliation of graphite in NMP (WJM–G). (b) Ultrasonic de-bundling of single/double walled carbon nanotubes (SDWCNTs) in NMP. (c) Sketch of the manufacturing of electrodes and EDLCs: the hybrid WJM–G/SDWCNTs dispersion prepared by mixing WJM–G and de-bundled SDWCNTs; the as-produced dispersion is deposited by vacuum filtration; the resulting peeled off film is the freestanding electrode, which is cut in pieces that are used for assembling an EDLC in coin-cell architecture.

Among CNTs, single/double-walled CNTs (SDWCNTs) were specifically selected with the aim to limit the active material costs compared to that of the single-walled counterpart[99] and to exploit their superior specific surface area (*i.e.*, theoretical specific capacitance) (~1315 $m^2$ $g^{-1}$ for single-walled CNTs,[100–102] ~800 $m^2$ $g^{-1}$ for double-walled CNTs[100]) compared to multi-walled CNTs (~50 $m^2$ $g^{-1}$ for 40-wall CNTs[100]). Experimentally, SDWCNTs were dispersed in NMP, de-bundled by an ultrasonication-based approach[103] and mixed with the WJM–G dispersion in a 1:1 material weight ratio. By taking advantage of the dimensional form of both graphene flakes (2D) and SDWCNTs (1D), EDLCs electrodes were produced by vacuum filtration of the as-produced hybrid dispersions through microporous nylon membranes. This allowed us to produce self-standing electrodes for EDLCs. The latter have been first optimized in reliable coin cell configuration using organic electrolyte (1 M tetraethylammonium tetrafluoroborate –TEABF$_4$– in propylene carbonate –PC–) and, subsequently, produced in flexible pouch-cell configuration using hydrogel-polymer electrolyte (poly(vinyl alcohol) –PVA– doped with H$_3$PO$_4$). By increasing the active material mass loading up ~30 mg cm$^{-2}$, the EDLCs reached a C$_{areal}$ up to ~317 mF cm$^{-2}$ (corresponding to an electrode C$_{areal}$ of ~634 mF cm$^{-2}$), an energy density of 539 μWh cm$^{-2}$ at 1.3 mW cm$^{-2}$, and operating power densities as high as 532 mW cm$^{-2}$. The EDCLs show excellent cycling stability and, in their flexible configuration, outstanding flexibility even under highly folded states (angles up to 180°).

Our approach represents a valuable method for the design/manufacturing of graphene-based electrodes for EDLCs with high active material mass loading allowing ultrahigh areal performance and flexibility to be achieved.

**Results and Discussion**

**Wet-jet milling graphene production and characterization**

The production of graphene flakes was carried out by means of a WJM approach, whose main steps are summarised in **Fig. 1**a.[81] Briefly, a dispersion of graphite in NMP flows in a pneumatic valve that pressurises this dispersion inside the processor, reaching high-pressures (180 – 250 MPa) when passing through a sequence of nozzles having different size (from 0.3 to 0.1 mm of diameter).[81] The use of high-pressure generates high shear forces that promote the exfoliation of the dispersed material.[81]. Following this procedure, by properly selecting a suitable nozzle diameter and size-sequence, it is possible to tailor-make the exfoliation process.[81] Specifically, we produced the graphene dispersion with an 8-passes procedure, detailed in the Experimental section. Compared to protocols previously reported in our WJM exfoliation studies,[81] our procedure increased the number of passes from 4 to 8 with the aim to further reduce the lateral dimension and thickness of the as-produced flakes, which is pivotal for obtaining high-specific surface active materials to be exploited in EDLCs.[1,2,4,5]

The morphology (*i.e.*, lateral size and thickness) of the exfoliated graphitic flakes was evaluated by transmission electron microscopy (TEM) (**Fig. 2**a-b) and atomic force microscopy (AFM) (**Fig. 2**c-d). The sample consisted of irregularly shaped (Figure 2a) and nm-thick flakes (Figure 2c). Statistical analysis indicates that the thickness and the lateral size of the flakes approximately follow log-normal distributions peaked at ~470 μm (Figure 2b) and ~1.67 nm (Figure 2d), respectively. These results indicate that the WJM–G flakes consist mainly of SLG/FLG flakes (SLG thickness ~0.34 nm[50]), in agreement with our previous works.[81] Raman spectroscopy measurements were performed to assess the structural properties of the WJM–G flakes (**Fig. 2**e-f). The Raman spectroscopy characterization, carried out at an excitation wavelength of 514 nm, focused on the four main spectral features exhibited by SLG/FLG: D (~1380 cm$^{-1}$)[104], G (~1585 cm$^{-1}$)[105,106], D' (~1620 cm$^{-1}$)[107] and 2D (~2700 cm$^{-1}$)[106] peaks (see Appendix A for a detailed analysis of SLG/FLG Raman spectrum). **Fig. 2**e shows a representative Raman spectrum of WJM–G. The analysis of the 2D peak reveals that it is composed of two contributions, namely $2D_1$ and $2D_2$,[108] which provide information about the number of layers.[106,109] Experimentally, the $2D_2$ component intensity in graphite is twice the one of $2D_1$.[106,109] In SLG, just a single sharp peak appears[108], while for FLG the $2D_1$ band is more intense than the $2D_2$ band.[106,109] Contrary, multi-layer (> 5 layers) graphene exhibits a 2D band which is almost equal to the one of graphite.[110,111] Based on these observations, the Raman spectrum of **Fig. 2**e can be attributed to FLG. The detailed evaluation of the intensity ratio $I(2D_2)/I(2D_1)$, full width at half maximum of G –FWHM (G)– and 2D components position is reported in Supporting Information (**Fig. S1**), pointing out that the WJM–G dispersion is mainly composed of a combination of SLG and FLG, in agreement with the AFM analysis. Moreover, the absence of any correlation between the intensities ratio $I(D)/I(G)$ and the FWHM(G) (**Fig. 2**f) indicates that defects of the WJM–G flakes were located at the flakes edges,[110,112,113] without affecting the structure of the basal planes.

**Preparation of active material dispersion**

The electrodes were produced by using a hybrid active material dispersion composed by WJM–G flakes and SDWCNTs. More in detail, the SDWCNTs dispersion in NMP was produced by de-bundling the powder through an ultrasonic bath-assisted method (**Fig. 1**b) (see Experimental section for further details).[103] Subsequently, SDWCNTs dispersion was mixed with WJM–G dispersion with a material weight ratio of 1:1, in agreement with previous optimized hybrid graphene/CNT[76] and graphite/CNT[114] EDLCs.

**Production and characterization of the electrodes and EDLCs**

The EDLC electrodes were produced by depositing the hybrid WJM–G/SDWCNT dispersions by vacuum filtration exploiting a nylon micro-pore membrane (**Fig. 1**c).

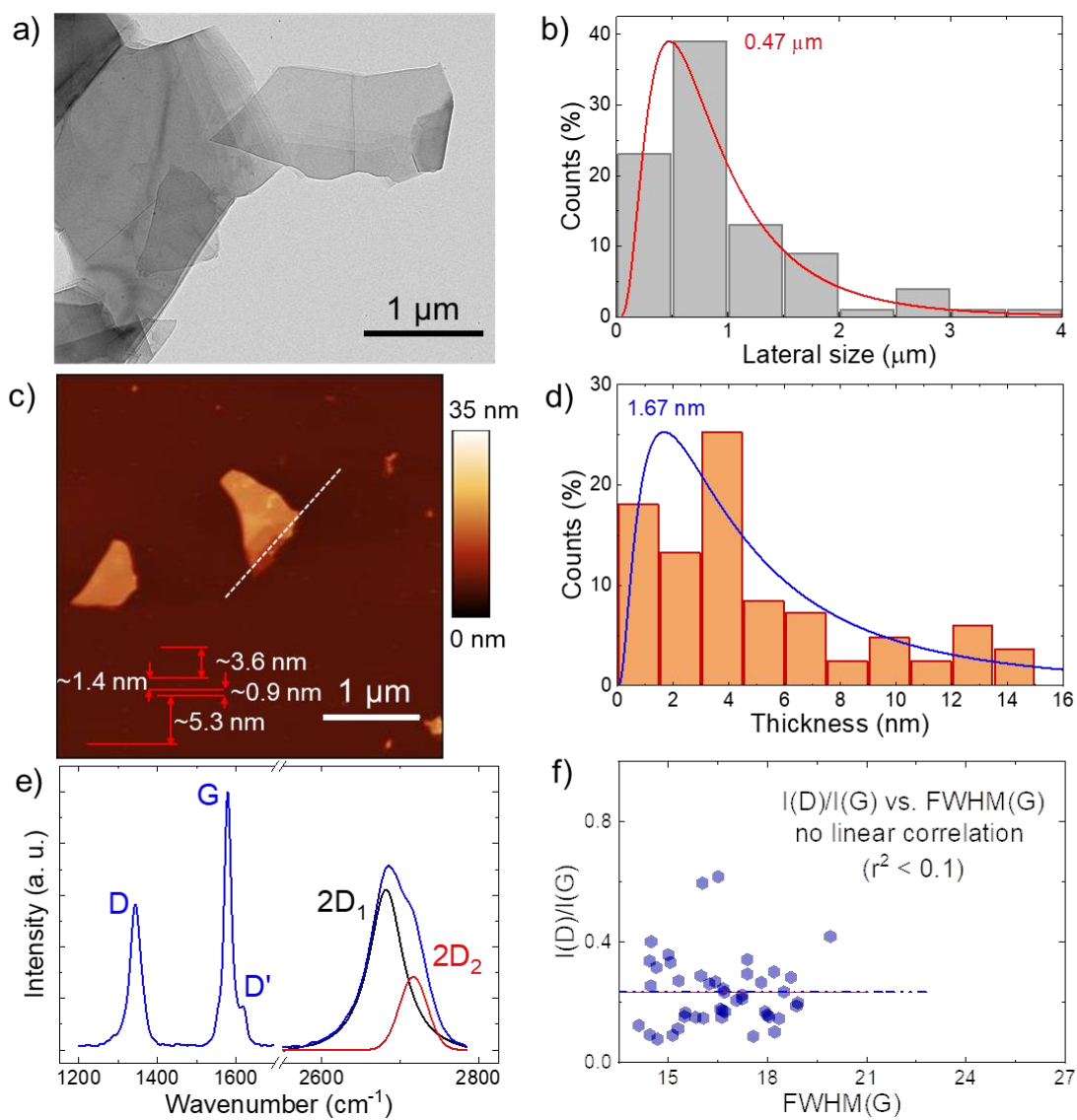

**Fig. 2.** Morphological and structural analysis of WJM–G. (a, b) Representative TEM image of WJM–G flakes (a) together with the corresponding statistical analysis of the lateral size (b), performed on 90 flakes. (c, d) Representative AFM image of WJM–G flakes (c) with the corresponding statistical analysis of the thickness, performed on 90 flakes. (e) Representative Raman spectrum of WJM–G normalized to the intensity of the G peak. The deconvolution of the 2D region is also shown: $2D_1$ peak (black), $2D_2$ peak (red). (f) I(D)/I(G) and the FWHM(G) plot calculated by Raman spectra on 50 flakes.

The WJM–G/SDWCNTs was washed by 2-propanol (IPA) (low-boiling point solvent) to remove residual NMP. There is a peeling off of the WJM–G/SDWCNT films from the nylon membrane during the drying process at room temperature, resulting in self-standing films used as EDLC electrodes. **Fig. 3**a shows a picture of a representative WJM–G/SDWCNTs electrode. By controlling the volume of the deposited dispersion, WJM–G/SDWCNT electrodes were produced with active material mass loading ranging from ~1.6 to ~30 mg cm$^{-2}$.

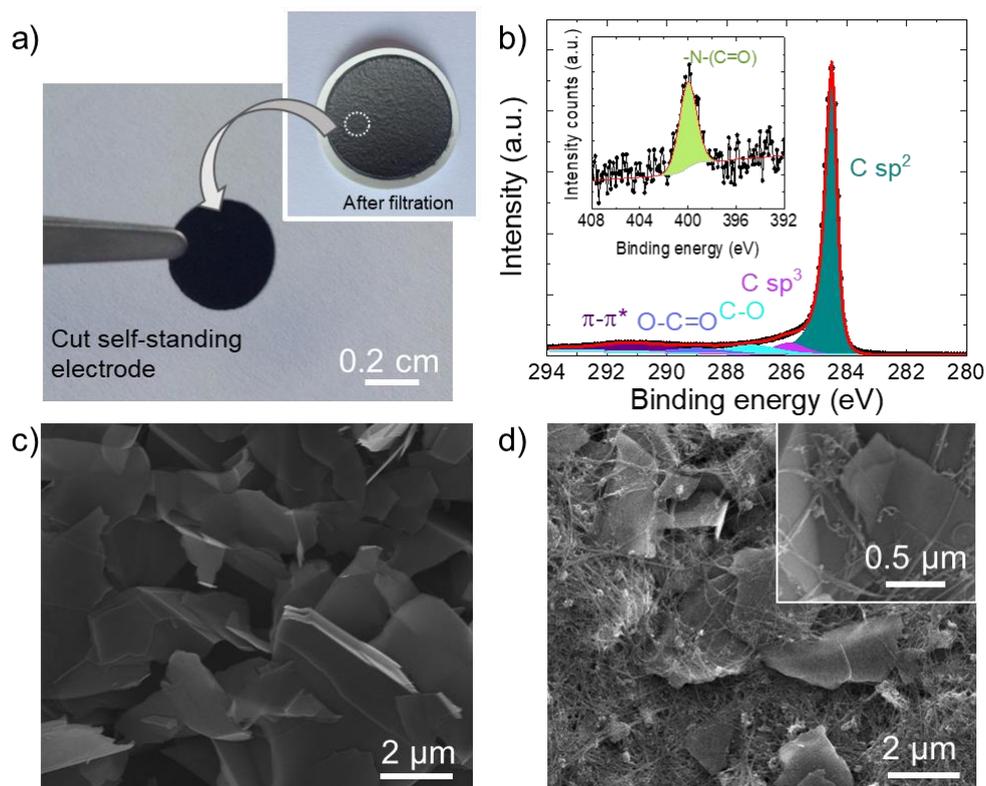

**Fig. 3.** Chemical and morphology characterization of the as-produced EDLC electrodes. (a) Photograph showing the deposited material on the membrane together with a representative picture of a self-standing WJM–G/SDWCNTs electrode. (b) C 1s and N 1s core-level (inset) XPS spectra of the WJM–G electrode, together with the corresponding deconvolution. (c, d) SEM images of representative WJM–G (c) and WJM–G/SDWCNTs (d) electrodes. The inset to panel (d) shows a high-magnification SEM image of the WJM–G/SDWCNTs electrode.

Electrodes based on WJM–G (average active mass loading of ~11.4 mg cm$^{-2}$) were also produced as reference. X-ray photoemission spectroscopy (XPS) measurements were acquired to ascertain the C atomic network and the chemical quality of the WJM–G films. The C 1s core-level XPS spectrum (**Fig. 3**b) shows an asymmetric peak-shape centred at 284.5 eV, corresponding to C sp$^2$, typical of graphitic-like compounds, together with the π-π* transition at 290.9 eV. Graphitic carbon accounts for 85.4% of the total carbon content of the sample. Defect- and oxygen-related components, *i.e.* C-O (at 285.9 eV), C=O (at 287.1 eV) and O-C=O (at 288.9 eV) marginally contribute to the spectrum (percentage content %c = 5.0%, 5.7% and 3.9%, respectively). The N 1s core-level XPS spectrum (inset of **Fig. 3**b) was acquired to evaluate the NMP residual into electrodes after IPA-based washing procedure. In fact, the band peaking at ~400.0 eV is associated to the amide groups (-N-(C=O)),[115,116] which are signature of the presence of NMP. However, the elemental analysis shows an overall %c of N ~0.1%, thus indicating that NMP was almost completely removed by IPA-based washing of the electrodes. The morphology of the electrodes was assessed by scanning electron microscopy (SEM) measurements. **Fig. 3**c shows a SEM image of a representative WJM–G electrode, displaying the laminar structure of the WJM–G flakes. **Fig. 3**d reports a SEM image of a representative hybrid electrode, in which the WJM–G flakes are wrapped and linked by SDWCNTs.

Symmetric EDLCs were produced by assembling two WJM–G/SDWCNTs electrodes with the same active material mass loading in coin cells configuration. The latter was chosen since it is a benchmark methodology providing a reliable assembly of EDLCs.[117] Briefly, coin cells were produced using CR2032 cases, a glass fibre separator and a 1 M TEABF$_4$ solution in PC as electrolyte (see the Experimental section for further details). Such electrolyte is widely used for both commercial and research-prototype due to its capability to operate without degradation in a wide range of voltages (>3 V) and temperatures (>50 °C).[118–120]

**Electrochemical characterization of the EDLCs**

The electrochemical performance of WJM–G/SDWCNTs EDLCs were evaluated by cyclic voltammetry (CV) (**Fig. 4**) and galvanostatic charge/discharge (CD) (**Fig. 5**) measurements.
The evaluation of the EDLCs areal capacitances ($C_{areal}$) was carried out from the CD measurements (**Fig. 5**) by using the equation $C_{areal} = (|i|\ t_d)/(A\ \Delta V)$.[121,122] In the previous equation, |i| is the module of the applied current (i), $t_d$ is the discharge time, $\Delta V$ is the voltage window of the measurements and A is the geometrical area of the electrodes. All the electrochemical characterizations were performed in the voltage window 0–3.5 V in order to avoid parasitic chemical reactions (breakdown voltage of TEABF$_4$/PC >3.5 V).[118] **Fig. 4**a shows the comparison between the CV curves of a hybrid WJM–G/SDWCNTs EDLC and a reference WJM–G EDLC with similar electrode mass loading (13.5 and 11.4 mg cm$^{-2}$, respectively). Both the CV curves show a nearly rectangular shape, characteristic of the electrochemical double-layer capacitor behaviour.[5,122,123] The area of the CV curve can be quantitatively correlated with the $C_{areal}$ of the EDLCs.[121] The area of the CV curve of the hybrid WJM–G/SDWCNTs EDLC increases by 1301% compared to that of WJM–G EDLC. This is ascribed to the presence of SDWCNTs as spacers between WJM–G flakes, which tend to restack in WJM–G EDLC [52,58,76,124] causing a decrease of the EDLC capacitance. **Fig. 4**b shows the CV curves of WJM–G/SDWCNTs EDLCs with various electrode active material mass loadings (ranging from 1.6 to 30.3 mg cm$^{-2}$) at a voltage scan rate of 100 mV s$^{-1}$. The current densities increase with increasing the electrode active material mass loading (so higher active material surface area is available), and consequently, the $C_{areal}$, which is quantitatively correlated to the area of the CV curves, rises.[121] Cyclic voltammetry measurements at voltage scan rates ranging from 0.01 to 50 V s$^{-1}$ were performed to assess the rate capability of the EDLCs. **Fig. 4**c,d shows such CV curves for the WJM–G/SDWCNTs EDLC with electrode active material mass loading of 13.5 mg cm$^{-2}$. Notably, the WJM–G/SDWCNTs EDLC still exhibits a capacitive behaviour even at scan rates up to 50 V s$^{-1}$. The biconvex lens-shape of the CV curve for the highest voltage scan rate can be ascribed to the equivalent series resistance of the EDLC at current densities higher than tens of mA cm$^{-2}$. This contribution is associated to the electrolyte resistance of the ions into the porous active material films,[122,125–128] as well as the electrical resistance of the latter.[122,125] **Fig. 5**a shows the comparison between the CD curves of WJM–G and WJM–G/SDWCNTs with similar electrode active material mass loading (11.4 and 13.5 mg cm$^{-2}$ respectively). Both the CD curves display nearly triangular shapes in the considered voltage window, confirming their electrochemical double-layer capacitor behaviour,[5,122,123] in agreement with CV analysis. Although the WJM–G/SDWCNTs EDLC was tested at a current density 18% higher than the one used for the WJM–G EDLC, it shows a CD curve that lasts >26-fold time that of the WJM–G-based one. This reflects the superior $C_{areal}$ (129.6 mF cm$^{-2}$) of the WJM–G/SDWCNTs EDLC compared to that of WJM–G EDLC (4.6 mF cm$^{-2}$), see **Fig. S2** for the $C_{areal}$ comparison. **Fig. 5**b reports the CD curves of the WJM–G/SDWCNTs EDLC with electrode active material mass loading of 13.5 mg cm$^{-2}$ at various current densities applied (between 0.3 and 6.8 mA cm$^{-2}$). The CD curves of the same electrode at applied current densities between 13.5 and 135.4 mA cm$^{-2}$ are reported in **Fig. S3**. **Fig. 5**c shows the comparison between the CD curves of the WJM–G/SDWCNTs EDLCs with various electrode active material mass loadings (ranging from 1.6 to 30.4 mg cm$^{-2}$) at comparable applied current densities.

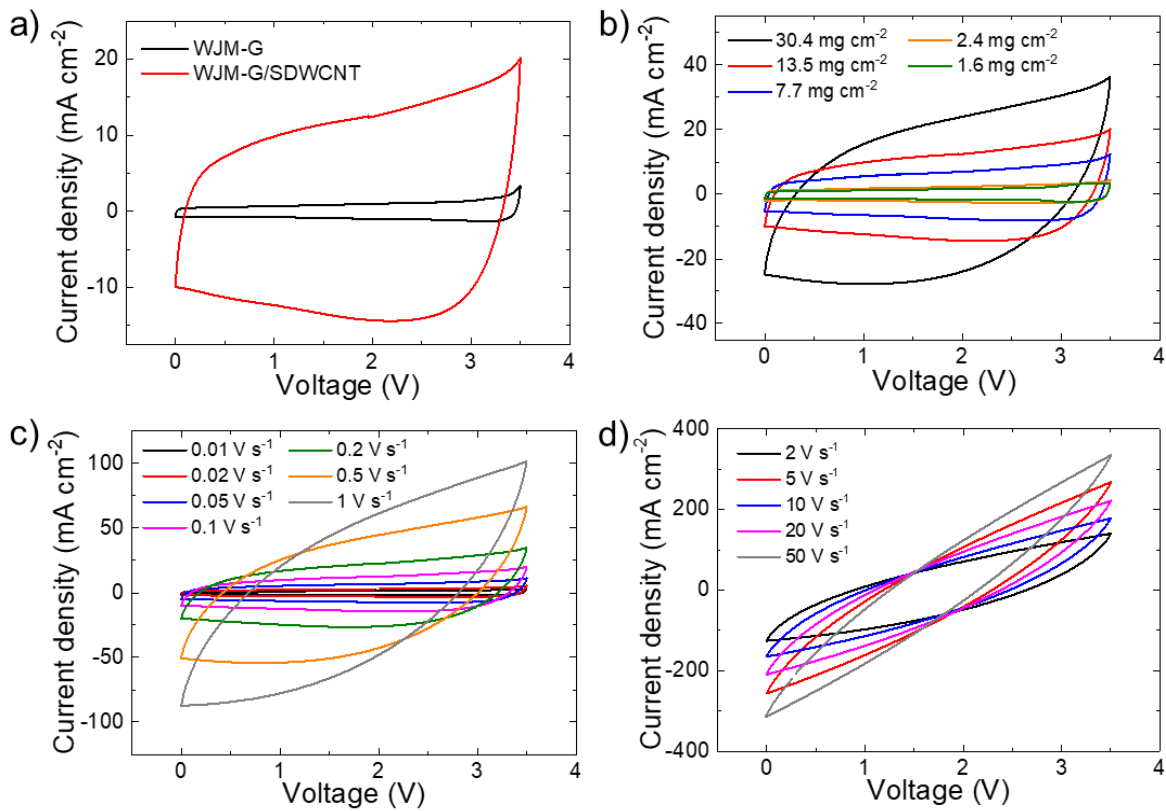

**Fig. 4.** Cyclic voltammetry characterization of the produced EDLCs. (a) Comparison between the CV curves of WJM–G EDLC (electrode active material mass loading of 11.4 mg cm$^{-2}$) and WJM–G/SDWCNTs EDLC (electrode active material mass loading of 13.5 mg cm$^{-2}$) at a voltage scan rate of 100 mV s$^{-1}$. (b) Comparison between the CV curves of WJM–G/SDWCNTs EDLCs with various electrode active material mass loadings (ranging from 1.6 to 30.3 mg cm$^{-2}$) at a voltage scan rate of 100 mV s$^{-1}$. (c, d) CV curves of the WJM–G/SDWCNTs EDLC with electrode active material mass loading of 13.5 mg cm$^{-2}$, at scan rates ranging from 0.01 V s$^{-1}$ to 1 V s$^{-1}$ (c) and from 2 to 50 V s$^{-1}$ (d).

These results demonstrate that the CD curves broaden in time as the mass loading increases, boosting the $C_{areal}$ from 15.6 mF cm$^{-2}$ to 313.5 mF cm$^{-2}$ for the EDLC with electrode active material mass loading of 1.6 mg cm$^{-2}$ and 30.4 mg cm$^{-2}$, respectively. **Fig. 5**d shows the dependence of the $C_{areal}$ of the WJM–G/SDWCNTs EDLCs on the applied current densities. The EDLCs having the electrode with the highest active mass loading (30.4 mg cm$^{-2}$) reached a $C_{areal}$ of 317 mF cm$^{-2}$ (*i.e.*, electrode $C_{areal}$ of 634 mF cm$^{-2}$) at a current density of 0.8 mA cm$^{-2}$. Interestingly, at a current density of 304.1 mA cm$^{-2}$, the EDLCs still exhibits a remarkable value of 7.1 mF cm$^{-2}$. Electrochemical impedance spectroscopy (EIS) measurements also confirm the increment of the $C_{areal}$ with the increase of the electrode active material mass loading (see Supporting Information for the detailed discussion, **Fig. S4**), in agreement with the CV measurements (Fig. 4b). Notably, the relationship between the $C_{areal}$ (as extrapolated by EIS data) and the electrode active material mass loading is almost linear (**Fig. S5**), indicating an effective electrolyte ion accessibility to the surface of the WJM-G/SDWCNTs electrodes, even in the case of the highest mass loading (> 30 mg cm$^{-2}$). As shown in **Fig. 5**e, the WJM–G/SDWCNTs EDLC shows excellent cycling stability (electrode active material mass loading of 13.5 mg cm$^{-2}$). In fact, stable electrode $C_{areal}$ of ~124 mF cm$^{-2}$ (EDLC energy density of ~106 μWh cm$^{-2}$) was measured over 10000 CD cycles at a current density of 7.7 mA cm$^{-2}$, corresponding to a EDLC power density of 235 mW cm$^{-2}$.

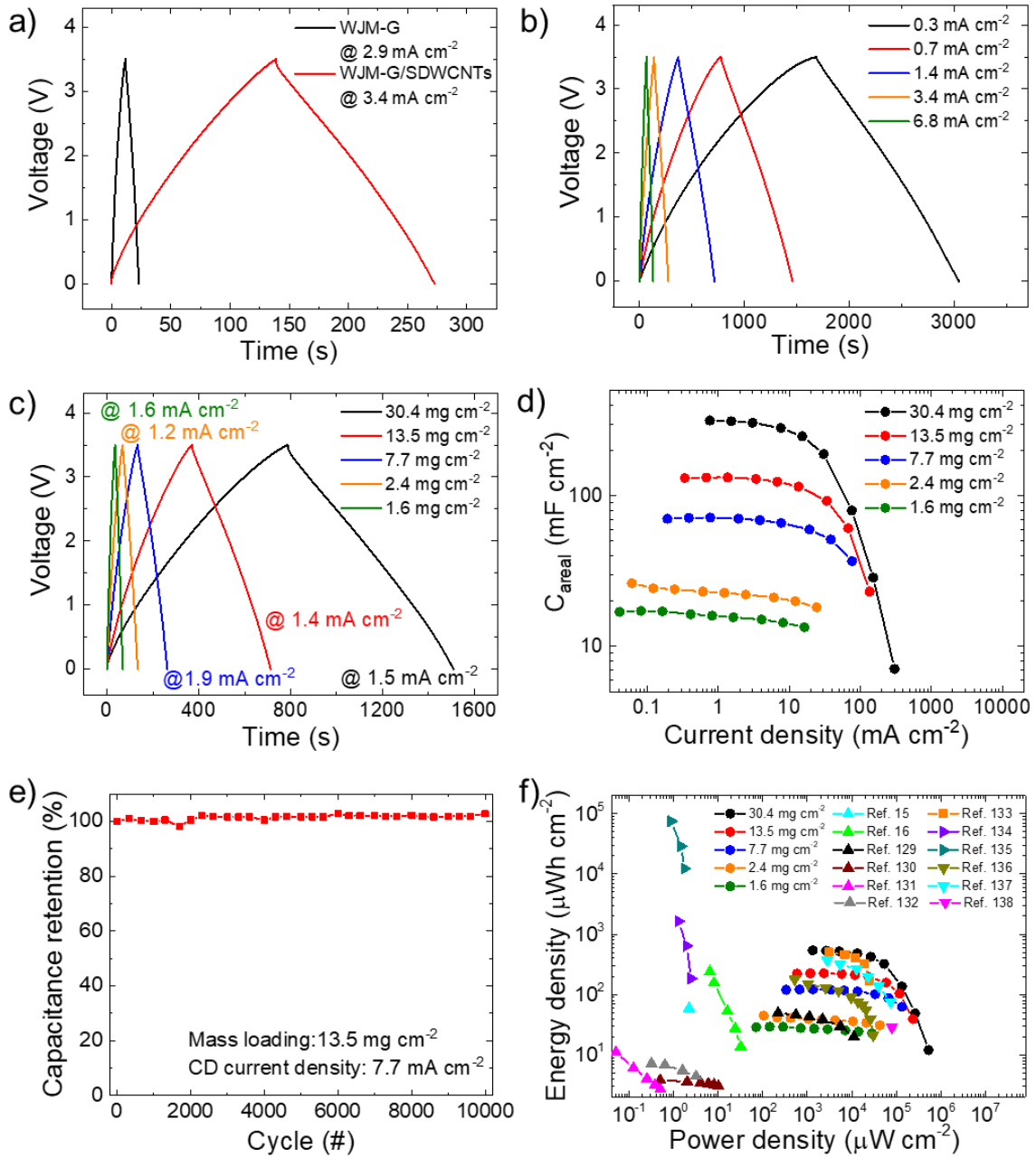

**Fig. 5.** Galvanostatic charge/discharge measurements of the as-produced EDLCs. (a) Comparison between the CD curves of the WJM–G EDLC (electrode active material mass loading of 11.4 mg cm$^{-2}$) and WJM–G/SDWCNTs EDLC (electrode active material mass loading of 13.5 mg cm$^{-2}$) at an applied current density of 2.9 mA cm$^{-2}$ and 3.4 mA cm$^{-2}$. (b) CD curves of the WJM–G/SDWCNTs EDLC with electrode active material mass loading of 13.5 mg cm$^{-2}$ electrode, at current densities ranging from 0.3 mA cm$^{-2}$ to 6.8 mA cm$^{-2}$. (c) Comparison between the CD curves of WJM–G/SDWCNTs EDLCs with different electrode active material mass loadings (ranging from 1.6 to 30.4 mg cm$^{-2}$), at comparable applied current densities. (d) Dependence of the calculated $C_{areal}$ on the applied current density for all the WJM–G/SDWCNTs EDLCs with different electrode active material mass loadings. (e) Capacitance retention of WJM–G/SDWCNTs EDLC with electrode active material mass loading of 13.5 mg cm$^{-2}$ over 10000 CD cycles at an applied current density of 7.7 mA cm$^{-2}$. (f) Comparison between the Ragone plot of the WJM–G/SDWCNTs EDLCs (for all investigated electrode active material mass loadings) and some relevant EDLCs reported in literature (▲: ref. [132], data calculated from reported thickness; ref. [15,16,129–132]. ■: ref. [133]. ▶: ref. [134,135]. ▼: ref. [136–138], data calculated from the reported mass loadings/thicknesses).

**Fig. 5**f shows the Ragone plots measured for the WJM–G/SDWCNTs EDLCs with various electrode active material mass loadings. Noteworthy, these EDLCs exhibit energy and power densities ranging from

28.8 to 539 µWh cm$^{-2}$ and from 71.4 µW cm$^{-2}$ to 532 mW cm$^{-2}$, respectively. These values are competitive with the ones reported in literature on EDLCs with relevant areal performances, including: aqueous electrolyte-based EDLCs [15,16,129–132], gel electrolyte-based EDLCs[133], porous silicon-based EDLCs [134,135] and graphene-based technologies [136–138]. Porous silicon-based EDLCs exhibit the highest energy densities among the reported results. However, the energy densities fade significantly with increasing the power densities (for example, in ref. [134] the starting energy density of ~1641 µWh cm$^{-2}$ at ~1.3 µWcm$^{-2}$ reduces to ~184.2 µWh cm$^{-2}$ at ~2.5 µW cm$^{-2}$). Similar trends are also observed in most of the other examples. Differently, WJM–G/SDWCNT EDLCs show a similar energy density value over more than three orders of magnitude of power densities. For example, the EDLCs with electrode active material mass loading of 30.4 mg cm$^{-2}$ exhibited an energy density of 539 µWh cm$^{-2}$ at 1.3 mW cm$^{-2}$ and 480 µWh cm$^{-2}$ at 13.3 mW cm$^{-2}$, which correspond to an energy retention of 89%. The EDLC adopting an electrode active material mass loading of 2.4 mg cm$^{-2}$ shows energy retention of 69% throughout power densities between 0.1 and 42.5 mW cm$^{-2}$. The extraordinary areal performance expressed by WJM–G/SDWCNT EDLCs can positively influence the EDLC packaging design and weight. In fact, in order to achieve the same areal performance of a single WJM–G/SDWCNTs EDLC reported by our work, many EDLCs reported in literature have to be connected in parallel, complicating the overall EDLC unit architectures. For example, the WJM–G/SDWCNTs EDLC, with electrode active material mass loading of 30.4 mg cm$^{-2}$, exhibits a maximal energy density of 539 µWh cm$^{-2}$. This means that 2 EDLCs from ref. [137] (367 µWh cm$^{-2}$), 3 EDLCs from ref. [16] (240 µWh cm$^{-2}$) and 19 EDLCs from ref. [138] (29 µWh cm$^{-2}$) are needed to achieve the same performance. This results in complicated and/or bulky architectures. This consideration is of paramount importance for wearable applications, in which the surface available from a human body and load are limited (~1.5 m$^2$)[14]. Thus, wearable devices should not be too big and/or heavy to prevent adding uncomfortable burdens on the body. To prove the applicability of WJM–G/SDWCNT EDLCs for wearable electronics, flexible EDLCs in form of pouch cells have also been fabricated. In this case, a hydrogel-polymer, *i.e.*, poly (vinyl alcohol) (PVA) doped with H$_3$PO$_4$ was used as a quasi-solid-state electrolyte in order to eliminate the use of bulky and rigid encapsulation material to prevent the hazardous leakage of liquid electrolyte.[139–141] Moreover, among the quasi-solid-state electrolytes, PVA/H$_3$PO$_4$ mixture has been recognized as one of the best one for graphene-based EDLCs.[142,143] Silicon-adhesive Kapton® tape and PVA/H$_3$PO$_4$-soaked polyvinylidene fluoride (PVDF) membrane were used as encapsulant and separator, respectively. The thickness of the whole quasi-solid-state WJM–G/SDWCNTs EDLC (electrode active material mass loading of 5.0 mg cm$^{-2}$), including the contribution of both the packaging materials and the electrolyte, was equal to ~0.4 mm. Large-area and stacked EDLCs were also fabricated[144] (**Fig. 6**a, specification in Appendix A). **Fig. 6**b shows the CV curve of a quasi-solid-state WJM–G/SDWCNTs EDLC at various voltage scan rate. The cell voltage window was limited to 0–0.8 V to avoid electrochemical degradation of the electrolyte, in agreement with previous studies.[142],[143] The current increases linearly with the voltage scan rate, showing a capacitive behaviour up to 2 V s$^{-1}$ (inset panel), while beyond this value the electrolyte resistance (~5 Ω) causes resistive electrochemical behaviour. **Fig. 6**c reports the CD curves at different current density, together with the corresponding C$_{areal}$ values (inset panel). At the lowest current density of 0.3 mA cm$^{-2}$, the EDLC shows a C$_{areal}$ up to 67.1 mF cm$^{-2}$ (electrode C$_{areal}$ of 134.2 mF cm$^{-2}$), corresponding to an energy density of 6 µWh cm$^{-2}$ at power density of 100 µW cm$^{-2}$. The capacitance behaviour is retained up to current density of 25 mA cm$^{-2}$ (C$_{areal}$ = 15.9 mF cm$^{-2}$). Beyond this current density, the electrolyte resistance causes voltage losses comparable or superior to the operating voltage range of the quasi-solid-state EDLCs (0.8 V).

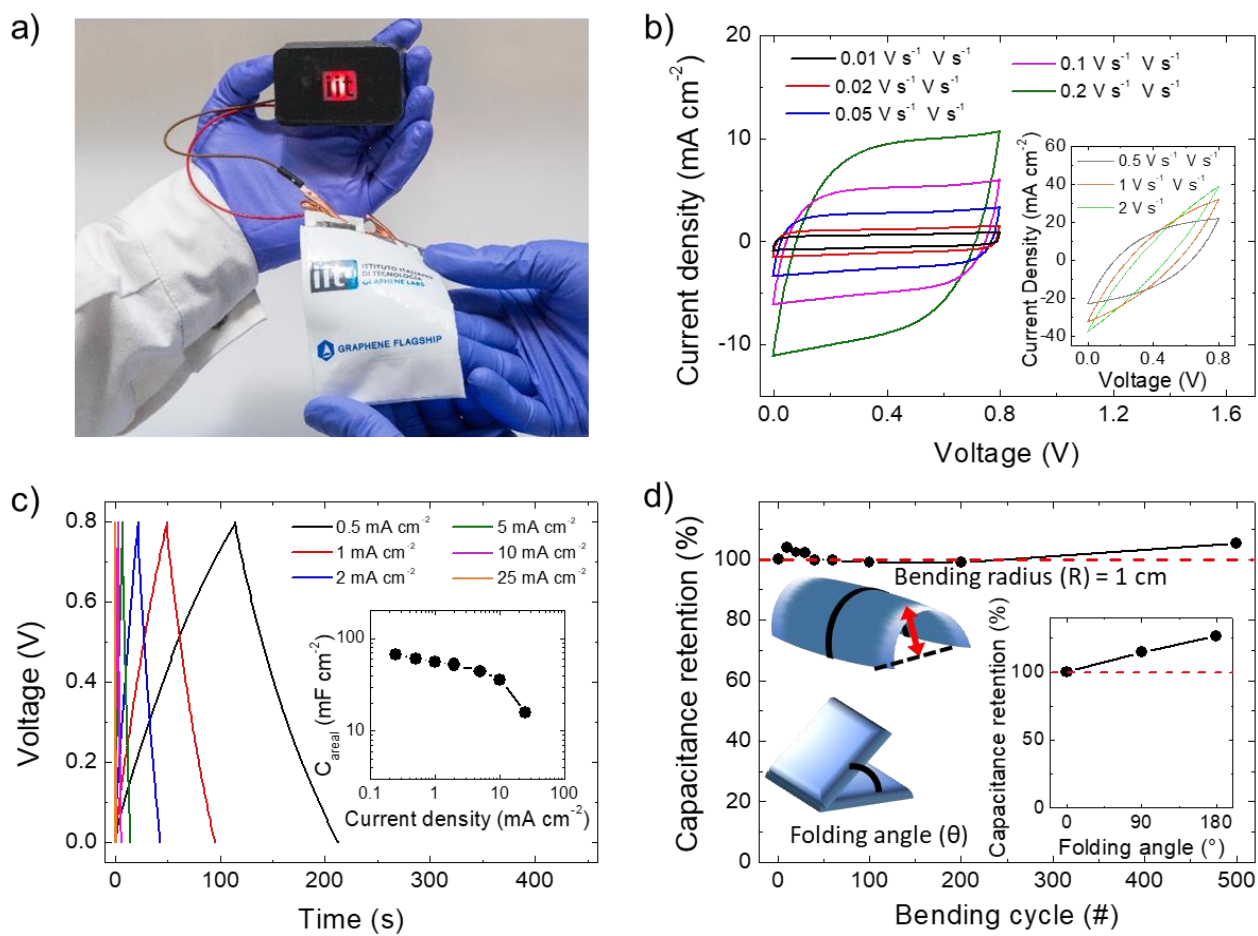

**Fig. 6.** Characterization of flexible quasi-solid-state WJM–G/SDWCNT EDLCs. (a) Digital photograph of a flexible quasi-solid-state WJM–G/SDWCNT EDCLs industrial prototype exhibited ad Mobile World Congress 2018 held in Barcelona. The electrodes were manually bended in order to show their mechanical flexibility. (b) CV curves of the WJM–G/SDWCNTs EDLC with electrode active material mass loading of 5.0 mg cm$^{-2}$, at scan rates ranging from 0.01 V s$^{-1}$ to 2 V s$^{-1}$. (c) CD curves of the WJM–G/SDWCNTs EDLC with electrode active material mass loading of 5.0 mg cm$^{-2}$ electrode, at current densities ranging from 0.5 mA cm$^{-2}$ to 25 mA cm$^{-2}$. The inset panel displays the dependence of the calculated C$_{areal}$ on the applied current density. (d) Capacitance retention of the quasi-solid-state WJM–G/SDWCNTs EDLC over 500 bending at R of 1. The inset panel shows the capacitance retention of folded quasi-solid-state WJM–G/SDWCNTs EDLC at θ of 0°, 90° and 180°.

Noteworthy, the $C_{areal}$ values reached by these quasi-solid state EDLCs with an electrode active material mass loading of 5.0 mg cm$^{-2}$ are comparable with those exhibited by coin-cell type EDLCs with a superior active material mass loading of 7.7 mg cm$^{-2}$. This result indicates that the electrolyte ion accessibility is also increased in the flexible EDLC configuration compared to the coin cell-type configuration. Interestingly, the whole flexible EDLC, including the packaging materials and the electrolyte, also exhibits a significant volumetric performance (*i.e.*, volumetric capacitance –$C_{vol}$– of ~1.75 F cm$^{-3}$ at 0.5 mA cm$^{-2}$, 1.1 F cm$^{-3}$ at 5 mA cm$^{-2}$).[63] **Fig. 6**d displays the capacitance retention plots over bending cycles adopting a bending radius (R) of 1 cm. The device shows an optimal capacitance retention over more than 500 bending cycles, showing an advantageous slight increase of the $C_{areal}$ (+6%). The latter could be ascribed to either a restrained evolution of the electrode films or a slight change of the working conditions (*e.g.*, a relative change of room temperature of ±10%). The devices were also tested at different folding states, corresponding to the folding angles (θ) of 0° (unbent states), 90° and 180 (inset panel). Interestingly, the capacitance of the device increased by 15% and 26% at θ of 90° and 180°, respectively. This behaviour can be ascribed to a favourable nano/mesoscopic reorganization of the morphology of the electrode films in presence of a compressive strain occurring during folding.[145]

**Conclusions**

In this work, we have reported the manufacturing and characterization of hybrid graphene/carbon nanotube electrochemical double-layer capacitors (EDLCs). Graphene flakes have been produced in form of liquid dispersions by a high-throughput ($Y_W$ ~100%, $t_{1g}$ = 2.55 min, $V_{1g}$ = 0.1 L) wet-jet milling (WJM) exfoliation of graphite in N-methyl-2-pyrrolidone (NMP). The dispersion was mixed with those of de-bundled commercial single-/double-walled carbon nanotubes (SDWCNTs) in a material weight ratio of 1:1. The SDWCNTs effectively acted as spacers between the WJM–produced graphene flakes (WJM–G), preventing the re-stacking of the latter during film deposition. Hybrid WJM–G/SDWCNT electrodes were obtained by depositing the as-produced dispersion through vacuum filtration on nylon membranes. 2-propanol-based washing removes NMP residuals, while subsequent drying of the electrode at room temperature causes the peel-off of the WJM–G/SDWCNT films, resulting in self-standing, binder-free electrodes. By controlling the volume of the deposited dispersion, electrodes with active material mass loading as high as ~30 mg cm$^{-2}$ have been demonstrated. Symmetric WJM–G/SDWCNT EDLCs were assembled in coin cells by using an organic electrolyte (1 M tetraethylammonium tetrafluoroborate in propylene carbonate), exhibiting an areal capacitance ($C_{areal}$) as high as ~317 mF cm$^{-2}$, an energy density of 539 μWh cm$^{-2}$ at 1.3 mW cm$^{-2}$, and outstanding rate-capability (power densities up to 532 mW cm$^{-2}$). By using polymer gel electrolyte, quasi-solid-state WJM–G/SDWCNT EDCLs have been fabricated, showing outstanding mechanical flexibility under bending cycling (500 cycles at bending radius of 1 cm) and folding (folding angle of 90° and 180°). These outcomes are competitive with respect to existing technologies developed for wearable electronics. The combination of industrial-like production of active materials, facile fabrication of solution-processed EDLCs, and ultrahigh areal performance of the obtained EDLCs are highly promising for novel advanced EDLCs design.

**Experimental section**

**Exfoliation of graphite**
The graphite was exfoliated thrugh the WJM method, whose details are reported in ref. [81]. A starting dispersion containing graphite (200 g, +100 mesh, Sigma Aldrich) and NMP (99%, 20 L, Sigma Aldrich) was mixed in a container with a mechanical stirrer (Eurostar digital Ika-Werke).The resulting mixture was firstly passed through the 0.30 mm diameter nozzle once, then all the processed sample was consecutively passed through the 0.20 mm nozzle (1 time), the 0.15 mm nozzle (1 time) and, finally, the 0.10 mm (5 times). The eight-WJM passed sample in NMP resulted in the WJM–G dispersion.

**De-bundling of SDWCNTs**
A commercial mixture of SDWCNTs (outer diameter 1–4 nm, length 5–30 µm, Cheap Tubes) was dispersed in NMP (99%, Sigma Aldrich®) and de-bundled by means of a sonic tip (Branson®, 3/16' tip).[146,147] Pulses of 4s on and 2s off were used, with vibration amplitude of 45%. An ice bath under the rosette was used to minimize heating effects during the sonication process.

**Atomic Force Microscopy characterization**
A NanoWizard III AFM system (JPK Instruments, Berlin) was used in intermittent contact mode for the AFM measurements. Cantilevers with a nominal tip diameter of 10 nm (PPP-NCHR probe, Nanosensors) and a drive frequency of ~320 kHz were used. The images were collected with a scan rate of 0.7 Hz and a working set point above 70% of the free oscillation. The images were collected on an area of 5×5 µm$^2$ and have a resolution of 512×512 pixels and the corresponding height profiles were analysed with the JPK Data Processing software (JPK Instruments). A statistical analysis on 90 flakes was carried out on both sets of measurements and the data fitted with a lognormal distribution, which is an established statistic function used for fragmented systems.[148] The statistical analysis was performed by using Origin 9.1 software. The AFM measurements were carried out on drop-casted 1:30 diluted WJM–Graphene dispersion in NMP onto freshly-cleaved mica sheets (G250-1, Agar Scientific Ltd.) and dried under vacuum.

**Transmission Electron Microscopy characterization**
Graphene dispersions were drop-casted on carbon-coated copper grids (Ted Pella®) in a dilution of 1:50 in NMP. The TEM images were acquired with a JEOL JEM-1011 transmission electron microscope, operating at an acceleration voltage of 100 kV. The ImageJ® software (NIH) was used for morphological analysis. Statistical analysis was performed with OriginPro 9.1 software.

**Scanning Electron Microscopy characterization**
The SEM measurements of the as-prepared electrodes were carried out with a Helios Nanolab® 600 DualBeam microscope (FEI Company), operating at 5 kV and 0.2 nA. For the cross-section imaging, the electrodes were cut with a scalpel.

**Raman Spectroscopy characterization**
Raman characterization was carried out with a Renishaw microRaman invia 1000 using a 50× objective (numerical aperture of 0.75), with an excitation wavelength of 532 nm and an incident power on the samples of 5 mW. The samples were obtained by drop casting the 1:30 diluted WJM–Graphene dispersion in NMP onto a Si wafer covered with 300 nm thermally grown $SiO_2$ (LDB Technologies Ltd.). OriginPro 2016 was used to perform the deconvolution and statistics (50 spectra collected for each sample).

**X-ray Photoelectron Spectroscopy characterization**
The XPS measurements were performed by using a Kratos Axis UltraDLD spectrometer on WJM–G powder samples prepared by filtration following the same procedure that for the fabrication of the electrodes. The XPS spectra were acquired using a monochromatic Al Kα source operating at 20 mA and 15 kV. High-resolution spectra of C 1s and N 1s peaks were collected at pass energy of 10 eV and energy step of 0.1 eV. Data analysis was performed out with CasaXPS software (version 2.3.17).

**Preparation of electrolytes**
For liquid organic electrolyte, $TEABF_4$ (>99%, Sigma Aldrich®) (2.17 g) was dispersed into 10 mL of propylene carbonate (anhydrous, 99.7 %, Sigma Aldrich®) in Ar glove-box. The mixture was left under stirring until it becomes clear (ca. 2h), obtaining the electrolyte solution (1 M). The electrolyte was stored in an Ar glove-box for further use. For quasi-solid-state hydrogel-polymer electrolyte, 1 g of $H_3PO_4$ and 1 g of PVA (molecular weight: 89 000–98 000, Sigma-Aldrich) were added to 10 mL of deionized water. The whole mixture was stirred and heated to 80 °C until the solution became clear, obtaining $H_3PO_4$-doped PVA-based hydrogel-polymer electrolyte.

**Electrodes fabrication and coin cells assembling**
The WJM–G ink was mixed with a dispersion of SDWCNTs. Since their concentrations were known (10 g $L^{-1}$ and 5 g $L^{-1}$ for graphene and SDWCNTs, respectively), volumes of these materials were chosen to achieve a 1:1 material weight ratio. The resulting hybrid WJM–G/SDWCNTs dispersions were vacuum filtrated through nylon membranes (0.2 μm pore size, 25 mm diameter, Sigma Aldrich®). To achieve various mass loadings (ranging from 1.6 mg $cm^{-2}$ to 30.4 mg $cm^{-2}$), different volumes of the hybrid material were filtrated. Additional washing of the electrodes in IPA was performed to remove NMP residuals from the deposited films. This procedure was carried out by filtrating, after the complete deposition of the film, a volume of IPA which was always larger than the volume of NMP used for the deposition (*e.g.* for 5 mL of NMP, 10 mL of IPA were used). The as-produced films (with 3.14 $cm^2$ area) spontaneously peeled off from the nylon membrane during the drying at room temperature, resulting in self-standing electrodes, which were cut with a diameter size of 0.4 cm. These electrodes were assembled in CR2032 type coin cells (stainless steel, 20 mm diameter and 3.2 mm thickness, MTI Corporation) using glass fibre separator (GF/D Whatman®). 150 μL of 1 M $TEABF_4$ in PC were used as electrolyte. The coin cells were electrochemically characterised after 2 h from their crimping in order to allow the electrolyte to diffuse throughout the electrodes.
Flexible quasi-solid-state WJM–G/SDWCNTs EDLCs were assembled by using self-standing active material films as electrode (area of 2.2 $cm^2$), $H_3PO_4$/PVA mixture-soaked PVDF membrane as both electrolyte and separator, and silicon-adhesive Kapton® tape (Dupont®, from Tesa) as encapsulant material.

**Electrochemical characterization**
The electrochemical performance of the EDLCs was evaluated by CV, galvanostatic CD and electrochemical impedance spectroscopy (EIS) measurements. All the analyses were performed by using a VMP3 electrochemical workstation (BioLogic Science Instruments) controlled *via* ECLab® software by a computer. Cyclic voltammetry and CD tests were carried out respectively at voltage scan rates between 0.01 and 50 V $s^{-1}$ and at current densities ranging from ~0.02 to ~152 mA $cm^{-2}$, depending on the device. The current densities (*J*) were calculated by using the equation:

$$J = \frac{|i|}{A_e}$$

where "|i|" is the absolute value of the applied current and "$A_e$" is the surface of one electrode. The $C_{areal}$ of the EDLCs (comprising two electrodes) was estimated by using the equation:

$$C_{areal} = \frac{|i| \cdot t_d}{A_e \cdot \Delta V}$$

where "$t_d$" is the discharge time of the CD curve and "$\Delta V$" is the potential window.
The $C_{vol}$ of the entire flexible quasi-solid-state EDLCs was calculated by:

$$C_{vol} = \frac{|i| \cdot t_d}{V_e \cdot \Delta V}$$

where "$V_e$" is the volume of the entire EDLCs, including packaging materials and electrolyte.
Energy density ($E_d$) and power densities ($P_d$) were estimated from:

$$E_d = \frac{1}{2} C_{areal} (\Delta V)^2$$

$$P_d = \frac{E_d}{t_d}$$

Electrochemical impedance spectroscopy measurements of the EDCLs were acquired in the 0.01 Hz - 200 kHz frequency range at 0 V with an AC amplitude of 0.02 V..

**Acknowledgements**

This project has received funding from the European Union's Horizon 2020 research and innovation program under grant agreement no.785219-GrapheneCore2. The authors also thank IIT Electron Microscopy and IIT Clean Room facilities for the access to carry out TEM and SEM/EDS measurements, respectively, and Dr. F. De Angelis (Plasmon Nanotechnologies) for the access to the Raman equipment.